\documentclass[aps,pra,superscriptaddress,amsmath,amssymb,preprintnumbers,twocolumn,floatfix,showpacs,showkeys,10pt]{revtex4-1}
\usepackage{amssymb}
\usepackage{epsfig}
\usepackage[caption=false]{subfig}

\begin{document}
\title{  Quantum correlations in
continuos-time quantum walks of two indistinguishable particles}
\author{Claudia~Benedetti }
\affiliation{ Dipartimento di
  Fisica, Universit\`{a} degli Studi di Milano, Via Celoria 16, Milano I-20133
  Italy}
\author{Fabrizio~Buscemi }
\email{fabrizio.buscemi@unimore.it} \affiliation{ARCES and Dipartimento di
  Elettronica, Informatica, e Sistemi, Universit\`{a} di Bologna,
  Viale Risorgimento 2, I-40136 Bologna, Italy} \author{Paolo~Bordone}\affiliation{ Dipartimento di
  Fisica, Universit\`{a} di Modena e Reggio Emilia, and Centro S3, CNR-Istituto di Nanoscienze,Via Campi 213/A,  Modena I-41125,
  Italy}
\begin{abstract}
We evaluate the  degree of quantum correlation between two fermions (bosons)
subject to  continuous time
quantum walks in a one-dimensional  ring lattice with periodic boundary conditions. In our approach,
no particle-particle interaction is considered. We show that the interference effects due to exchange
symmetry  can result into the appearance of non-classical correlations.  The role played onto the appearance of quantum correlations  by  the quantum  
statistics of the particles, the boundary conditions, and the partition of the system
 is widely investigated. Quantum correlations also been investigated
in a model mimicking  the  ballistic evolution of two indistinguishable particles in a 1D continuous space structure.
Our results are consistent with recent quantum optics and electron quantum optics experiments
where the showing up of two-particle non-classical correlations  has been observed even in the absence of mutual
interaction between the particles.

\end{abstract}
\pacs{05.40.-a, 03.65.Ud, 03.67.-a}
\maketitle

\section{Introduction}  Quantum walks (QW) describe the random walk 
behavior of a quantum particle~\cite{dtqw}. Due to quantum mechanics effects,
such as the coherent superposition of wavefunctions and interference, QW 
exhibit a qualitatively different behavior  with respect to classical random
walks, such as the ballistic propagation of the wavefunction instead 
of the diffusive behavior exhibited by the classical probability distribution~\cite{farhi}.
Quantum effects make   QW extremely promising 
for the implementation of more efficient and faster research algorithms 
than the protocols commonly adopted in the classical computation~\cite{ambainis,shenvi,childs}.
Due to their potential application in quantum information science,
simple models of QW have been investigated and physically
implemented in various physical systems ranging from quantum optics~\cite{Bromberg, Rai, Peruzzo} to
nuclear magnetic resonance setups~\cite{Ryan,du}.

Two kinds of QW are considered in the literature: discrete-time quantum coined walks
and continuous-time quantum walks (CTQW). In the former, a two-level state,
the so-called coin, rules the unitary discrete-time  evolution of a particle
moving in a  lattice of sites. On the other hand, for  CTQW the evolution
of the particle is continuous in  time  and it is only determined by a Hamiltonian
whose terms represent transitions among the  lattice sites. CTQW have 
 successfully been implemented in lattice waveguide systems~\cite{Bromberg, Rai, Peruzzo},
where the appearance of non-classical correlations has been observed 
for  two-photon input states~\cite{Bromberg, Peruzzo}.
%
Indeed, due to indistinguishability
of the photons, probability amplitudes of two-particle wavefunctions can 
interfere, thus leading  to the formation of correlations  with no particle-particle  interaction.
Interference effects have also been successfully detected in the so-called electron quantum optics
experiments, namely quantum optics experiments-like with electrons in solid
state systems, where many-particle effects play a key role~\cite{Hei,Neder}.  Recent developments
in nanofabrication technology have allowed  exploiting the ballistic
electron transport along chiral  edge states at integer quantum Hall regime
for the experimental realization of Mach-Zehnder and Hanbury Brown-Twiss
interferometers~\cite{Hei,Neder}. In these studies, two-particle Aharanov-Bohm  oscillations, due to exchange symmetry,
have been related to the degree of quantum correlation  between the charge carriers,
even if electron-electron interaction is not taken into account~\cite{Samue}.

In the last decade, the notion of  quantum correlations (QC)
as a fundamental source  for the implementation of quantum information
algorithm and commonly referred as ÔÕentanglementÕÕ
has  been widely investigated in systems of identical
particles ~\cite{schliemann,zanardi,wiseman,Gittings, Dow}.  Indeed, its quantification is
certainly  crucial for  understanding  a number of physical phenomena involving correlated 
indistinguishable subsystems. The main difficulties appearing in the definition of a criterion able 
to classify and quantify the amount of QC among indistinguishable particles are closely related to exchange simmetry
which requires the symmetrization or the antisymmetrization of the quantum wavefunctions
describing bosons or fermions, respectively. Different approaches have been used to estimate
the degree of non-classical correlation in bipartite systems of identical particles~\cite{schliemann,zanardi,wiseman}. The Schliemann criterion relies on 
the fermionic analogous of the Schmidt decomposition, namely the Slater decomposition~\cite{schliemann}.
In the  approach developed by Zanardi~\cite{zanardi}, the entanglement is evaluated in terms
of the QC between modes 
by mapping the Fock space of the modes themselves 
into qubit states.  In the criterion proposed by Wiseman and Vaccaro~\cite{wiseman,Dow},
the entanglement of the particles is a sort of accessible entanglement,
i.e. the maximum amount of non-classical correlations which can be extracted
from the system by means of local operations and then transferred into conventional
quantum registers.

 While the correlation  between position and coin degrees of freedom of particles 
 has been widely investigated in discrete-time quantum walks~\cite{dtqw, Path, Berry, Sand}, 
 an exhaustive  analysis of  the building up of the QC in CTQW
 of identical particles is still lacking. The appearance of non-classical correlations in
  two-particle quantum optics  CTQW  setups
 has  been discussed only qualitatively~\cite{Bromberg,Peruzzo}.
 The aim of this work is to provide an accurate analysis of the  
 quantum correlation created in
 the CTQWs of two identical particles (both fermions and bosons)  
 in a one-dimensional (1D) system.  Specifically,  we will first examine
 the diffusion of two  fermions (bosons) in a  lattice of sites with periodic boundary conditions,
 in order to study the development of non-classical correlations in simple models 
 mimicking experimental quantum optics and electron quantum optics 
 setups. Then, we will analyze  such a system 
 in the limit of a large number of  lattice sites, with a vanishing 
 intersite distance, thus resulting into  the free propagation of  two
 identical particles along a 1D structure.

 Our model  does not take into account any  kind of particle-particle interaction,
 so that  the time evolution of the two-particle quantum state 
 involving exchange symmetry is essentially ruled by single-particle Hamiltonians describing
 CTQW. In this way,  the appearance of QC is only
 related to two-fermion (-boson) amplitudes interference due to the quantum statistics
 of the particles involved in the process. In this work,
 in order to estimate the degree of non-classical correlation between the positions of the particles,
 we adopt the criterion proposed by Wiseman and Vaccaro~\cite{wiseman, Dow}. In fact, 
 such a criterion, unlike the approach developed by Schliemann~\cite{schliemann},  behaves correctly under one-site (local) and two-site 
 (non local) transformations. Furthermore,
 its use does not lead to the violation of  the local number of particle superselection rule,
 as  may happen in the case of the procedure proposed by Zanardi~\cite{zanardi}.

The  paper is organized as follows. In Sec.~\ref{review},  we describe the theoretical approach 
used to quantify the QC between two indistinguishable particles. In Sec.~\ref{resultsI},
we study the CTQW of two fermions (bosons) in a lattice of sites with periodic boundary conditions and then
estimate the time evolution of the amount of QC of the system. In Sec.~\ref{resultsII},
we numerically evaluate the degree of non classical correlation between two identical particles free propagating in
a 1D structure. Finally, conclusions and discussions are given in Sec.~\ref{conclusions}.

\section{Evaluator of the degree of quantum correlations in bipartite systems of
identical particles}\label{review}
Here, we briefly illustrate  the  theorethical criterion adopted to evaluate  the degree of quantum
correlation in two-fermion and
-boson systems. It is based on the notion of entanglement of particles
proposed by Wiseman and Vaccaro~\cite{wiseman, Dow}.

By using the mode-occupation representation, an arbitrary
pure two-fermion (boson) state in a $M$-mode system can be expressed as
\begin{equation}
 |\Psi\rangle= \sum_{\{n\}} c_{\{n\}} |{\{n\}} \rangle,
\end{equation}
where  the integers $n_i$ of the set ${\{n\}} = n_1,\ldots,n_i,\ldots,n_M$
satisfy the relation $n_1+\ldots +n_i+\ldots+n_M=2$. Here, the ket $|{\{n\}} \rangle$
indicates the state vector in the Fock space with $n_i$'s particles in the $i$-th mode,
and the  $c_{\{n\}}$'s are the coefficients of  the linear superposition. While for bosons
$n_i$'s range from 0 to 2, for fermions the occupations numbers are restricted to be
0 or 1 due to the Pauli exclusion principle. 

As argued by some authors~\cite{zanardi},
a formal equivalence between the space of the occupation-number states and the tensor
product space of the modes can be established. In this way, the occupation number 
of each mode constitutes a distinct state of the mode itself.  In the Zanardi approach~\cite{zanardi},
the amount of non-classical correlations between the occupation numbers of the modes
controlled by two parties of the system, namely Alice and Bob, represents the so-called entanglement of modes. 
The latter does not always constitute a valid measure of  the true degree of quantum
correlation between Alice and Bob~\cite{Dow}. 
As a matter of fact, not only the entanglement of modes 
can give values different from zero even when applied to suitable single-particle states~\cite{bus2},
but also it does not take into account the local-particle number superselection rule (LPNSR)\cite{wiseman}.
Indeed to fully exploit the QC  between modes, 
Alice and Bob, at least in principle, must be able to arbitrarily measure and manipulate their
local systems. Unless each party of the systems possesses a definite number of particles,
this will lead to  a violation of  the LPNSR.

The Wiseman and Vaccaro criterion satisfies the LPNSR.  In such an
approach,  in addition to the two identical particles shared by  Alice
and Bob,  their quantum state $|\Psi\rangle$ involve a standard quantum register, namely a set of
distinguishable qubits. The entanglement of the particles $E_P$ is defined
as the maximum amount of QC that Alice and
Bob can produce between their standard quantum registers
by means of local operations. As a consequence of the LPNSR,
$E_P$, in place of the QC between the modes
that Alice and Bob have access to, is given by 
\begin{equation}\label{entuno}
 E_P(|\Psi\rangle)=\sum_{n=0}^2P_nE_M(|\Psi^{(n)}\rangle),
\end{equation}
where $|\Psi^{(n)}\rangle$ is the projection of the  quantum state $|\Psi\rangle$,
describing the global system,  onto the Fock subspace where Alice controls $n$ particles 
and Bob the remaining $(2-n)$ ones. $P_n=\langle\Psi^{(n)}|\Psi^{(n)}\rangle$
is the probability for Alice (Bob) of finding $n$ $(2-n)$ particles as a consequence of a measure of the local number
of particles, while
$E_M$ is the degree
of quantum correlation between the two sets of modes, each controlled by a party of the system.
In other terms, $E_P$ is the weighted sum of the entanglement
of  modes when the local particle number is measured. It is worth noting that $E_P$ depends upon
the partition of the system, that is  upon which  modes Alice and Bob control. From this point
of view, different partitions of the system can lead to different values of $E_P$.

For  two-particle pure states, the expression given in Eq.~(\ref{entuno}) takes
a simple form.  Indeed, only some  quantum states $|\Psi^{(n)}\rangle$, belonging to the Fock subspace
with a fixed local number of particles, give a non vanishing contribution to entanglement. Specifically,
let us consider the case of the set of states  $|\Psi^{(0)}\rangle $
where the local number  of particles possessed by Alice  is zero. 
Any state of such a set  can be written, in the mode occupation number, as
$|0_{n_A} \rangle \otimes |2_{n_B} \rangle$ and it is separable, that is it can be factorized
in a term describing Alice with no particle and in a term describing Bob with 
two particles. This implies that the contribution of  $|\Psi^{(0)}\rangle$
to $E_P$ is zero. Analogously, 
 the amount of quantum correlations stemming from set of states  $|\Psi^{(2)}\rangle $ is also zero.
This implies that  the bipartite entanglement is non-vanishing  only when both Alice and Bob have one particle, that is only for the quantum states belonging to
the set  $|\Psi^{(1)}\rangle$.
As shown in Ref.~\cite{wiseman}, $E_P$ becomes
 \begin{equation}\label{entdue}
 E_P=P_1 \epsilon[\rho_A^{(1)}],
\end{equation}
 where  $\epsilon$ indicates an arbitrary quantum binary entropy and 
  \begin{equation}\label{enttre}
 [\rho_A^{(1)}]_{kk^{\prime}}=\frac{\langle\Psi^{(1)}|a_k^{\dagger}a_{k^{\prime}}|\Psi^{(1)}\rangle}{\sum_{l}\langle \Psi^{(1)}|a_l^{\dagger}a_l|\Psi^{(1)}\rangle}
 \end{equation}
 is the single-particle matrix  describing the subsystem  controlled by Alice. The latter is
 obtained from $|\Psi^{(1)}\rangle$ by means  of the creation (annihilation) operators $a_k^{\dagger}(a_k)$ acting
 on Alice's modes only. The form of $E_P$ given in Eq.~(\ref{entdue}) holds for either bosons or fermions.

\section{Two-particle continuous-time quantum walks  on 1D ring lattices}\label{resultsI}
In this section,  a model  of two-fermion and two-boson CTQW in 1D graphs is 
solved by means of analytical and numerical techniques for a small and large number of nodes, respectively.  
The  time evolution  of the wavefunction describing the system is then used to
estimate the  degree of quantum correlation  according to the criterion given in Eq.~(\ref{entdue}).

Here we examine CTQW on a 1D ring lattices of $N$ sites
(with $N$ even) with periodic boundary conditions.
In agreement with previous single-particle investigations~\cite{blumen},
the topology of the graphs here considered is simple in the sense that each node is connected
to its two first neighbours.  Even if more complex networks,  among which two-dimensional lattices~\cite{Volta} or graphs with
larger connectivity~\cite{Xu}, could be examined,  our model  is good enough to describe experimental implementations of  CTQW,
such as  two-photon transport in an array of  waveguide lattices where non-classical correlations appear.
Therefore, though simple, it represents a valid tool to analyze the amount of QC appearing in two-particle CTQW.

The  two-particle Hamiltonian describing the dynamical evolution of  the system  
is given by
\begin{equation} \label{des}
 \mathcal{H} =\mathcal{H}^0_{\alpha}+\mathcal{H}^0_{\beta},
\end{equation}
where $\mathcal{H}^0_{\alpha(\beta)}$ is the single-particle Hamiltonian  acting on   the particle $\alpha(\beta)$:
\begin{equation} \label{single}
\mathcal{H}^0_{\alpha(\beta)} |j\rangle_{\alpha(\beta)}= \gamma \left(2|j\rangle_{\alpha(\beta)} -|{j-1}\rangle_{\alpha(\beta)}-|{j+1}\rangle_{\alpha(\beta)}\right).
\end{equation}
$|j\rangle_{\alpha(\beta)}$ indicates  the quantum state describing the particle $\alpha (\beta)$ localized
in the $j$ node and forming a complete, orthonormalized basis set, which span the 
whole accessible Hilbert space. $\gamma$ denotes the intersite transmission rate. Due to the periodic
boundary conditions, here we assume that the node $N+1$ coincides with node 1.
 Eq.~(\ref{single}) is the discrete
version of the Laplacian, and, in turns, the discrete version of the
Hamiltonian describing the free propagation of a particle in a lattice.

Given the form of the two-particle Hamiltonian in Eq.~(\ref{des}),  single-particle
dynamics allows one to estimate the time evolution of the two-fermion and -boson system
in the basis states  given by 
\begin{equation}  \label{fermion}
|jk\rangle_f=\frac{1}{\sqrt{2}}\left( |jk\rangle_{\alpha \beta} -|kj\rangle_{\alpha \beta}\right),
\end{equation}
and
\begin{equation} \label{boson}
|jk\rangle_b=\left\{\begin{array}{ll} \frac{1}{\sqrt{2}}\left( |jk\rangle_{\alpha \beta} +|kj\rangle_{\alpha \beta}\right): & j\neq k, \\
|jj\rangle_{\alpha \beta}: & j= k ,\end{array} \right .
\end{equation}
respectively.  Specifically in order to solve the model, which can be interpreted
as the free propagation of  two  identical particles in a periodic system,
we adopt the Bloch function approach, in analogy with the method
commonly used in solid-state physics~\cite{blumen}. The single-particle
Bloch states 
\begin{equation} 
|\phi_n\rangle=\frac{1}{\sqrt{N}} \sum_{j=1}^N \exp{\left(-i \frac{2\pi n}{N} j\right)} |j\rangle,
\end{equation}
are eigenstates of the Hamiltonian of  the  Eq.~(\ref{single}) with eigenvalues $E_n=2\gamma\left(1-\cos{\frac{2\pi n}{N} }\right)$,
and can be used to evaluate the coefficients describing the 
transition amplitude $\lambda_{k,j}(t)=\langle  k|e^{-iH_0t}| j\rangle$ from the state $|j\rangle$ at $t=0$ to state $|k\rangle$ at time $t$.
Indeed the latter can be expressed as:
\begin{widetext}
\begin{eqnarray} \label{singpart}
\lambda_{k,j}(t)&=&\sum_{n=1}^N \sum_{m=1}^N \langle  k |\phi_n\rangle \langle \phi_n|e^{-iH_0t}|\phi_m\rangle \langle \phi_m| j\rangle \nonumber \\
&=&\frac{\exp{\left(-2i\gamma t\right)}}{N}\sum_{n=1}^N\exp{\left(2i\gamma t\cos\left(\frac{2n\pi}{N}\right)\right)}\exp{\left(-i\frac{2\pi n}{N}(k-j)\right)}\label{beta},
\end{eqnarray}
\end{widetext}
where  is set $\hbar=1$.

By using the two-particle basis states written in Eqs.~(\ref{boson}) and (\ref{fermion}),
we can evaluate, in terms of the above single-particle coefficients,  
the two-boson (-fermion) transition amplitude ${\mu^{f(b)}}_{ks,jr}(t)$, from the state 
$|jr\rangle_{f(b)}$ at $t=0$ to  state $|ks\rangle_{f(b)}$ at time $t$. It reads:
\begin{equation}  \label{fer2}
{\mu^{f}}_{ks,jr}(t)= \lambda_{k,j}(t) \lambda_{s,r}(t)-\lambda_{k,r}(t) \lambda_{s,j}(t),
\end{equation}
and
\begin{widetext}
\begin{equation} \label{boso2}
{\mu^{b}}_{ks,jr}(t)=\left\{\begin{array}{ll}  \lambda_{k,j}(t) \lambda_{s,r}(t)+\lambda_{k,r}(t) \lambda_{s,j}(t)& k\neq s \quad \textrm{and} \quad j\neq r\\
\sqrt{2}\lambda_{k,j}(t) \lambda_{s,r}(t) & k= s \quad \textrm{xor} \quad j= r\\
\lambda_{k,j}(t) \lambda_{s,r}(t) & k= s \quad \textrm{and} \quad j= r\\
 \end{array} \right ..
\end{equation}
\end{widetext}
Once the time evolution of the two-particle quantum state 
$|jr\rangle_{f(b)}$ describing two fermions (bosons) initially localized
in the node $j$ and $r$ is known, the dynamics of the built up of the amount of quantum correlation
 can be quantified by using Eq.~(\ref{entdue}). To this aim,
different partitions of the system can be considered. 
Note that here the modes of the system
correspond to the sites of the lattice. In the simplest case,
Alice controls the first half of the lattice and Bob the second one,
namely $A=\{1, \dots, \frac{N}{2}\}$ and $B=\{\frac{N}{2}+1, \dots, N\}$. Thus,
the single-particle density matrix  of the Alice's subsystem given in Eq.~(\ref{enttre}) takes, for the initial state $|jr\rangle_{b(f)}$,  the form
\begin{equation} \label{matrice}
\left[{\rho(t)}_A^{(1)}\right]_{kk^{\prime}}=\frac{\displaystyle{\;\;\sum_{s=\frac{N}{2}+1}^{N}\mu_{ks,jr}^{\ast}(t)\mu_{k^{\prime}s,jr}(t)\;\;}}{\displaystyle{\;\;\sum_{k=1}^{\frac{N}{2}}\sum_{s=\frac{N}{2}+1}^N|\mu_{ks,jr}(t)|^2}}.
\end{equation}
Its quantum entropy can be quantified by means of the von Neumann entropy 
\begin{equation} \label{vonNeu2}
\epsilon_{vN}(t)=-\textrm{Tr}\left\{\left[{\rho(t)}_A^{(1)}\right] \ln{\left[{\rho(t)}_A^{(1)}\right]}\right\},
\end{equation}
which represents an appropriate evaluator of the degree of the correlation
in standard bipartite systems, and this  can be used in Eq.~(\ref{entdue}) in order to estimate of the entanglement of particles.
For sake of completeness, in the following, other possible  partitions of the system will  be examined,
where Alice and Bob can access to non-adjacent modes. The evaluation of $E_P$ in such cases requires to calculate again both the single-particle
density matrix of the subsystems and the probability of finding one particle
in each subsystem, being  the latter closely dependent upon the specific partition. 

\subsection{Small number of nodes}
Here we study, by means of analytical techniques, the quantum correlation dynamics
for the case of an 1D ring lattice composed of a small number of sites. Specifically, we take $N=4$.
 Various initial configurations of the  bosonic (fermionic)
system, that is the  states describing at time $t$=0 the two identical particles, should be examined.
However,  due to the periodic boundary conditions of the lattice under investigation,
the number of  initial setups of the system resulting into different time evolutions
of the QC turns out to be  smaller than the dimension of the Fock space of  the two-fermion (-boson) system.

\begin{figure}[htpb]
  \begin{center}
 \includegraphics*[width=0.6\linewidth]{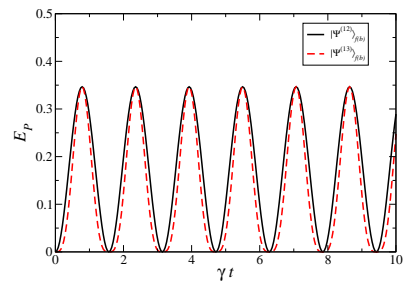}
    \caption{\label{fig1} (Color online) Time evolution of $E_P$ for  three input states, namely  $|12\rangle_{f(b)}$ (solid line), $|13\rangle_{f(b)}$ (dashed line),with Alice controlling sites $\{1,2\}$ and Bob $\{3,4\}$.  For sake of clarity, the entanglement dynamics of the initial state $|23\rangle_{f(b)}$  has not been reported being always null.}
 \end{center}
 \end{figure}

First, let us analyze the case of particles initially localized in two different nodes for the partition $A=\{1, 2\}$ and $B=\{3,4\}$. Three 
configurations are possible: both particles occupy the Alice (Bob) modes, that is the quantum state
$|12\rangle_{f(b)}$ ($|34\rangle_{f(b)}$);  one particle is in an Alice site  and the other one is localized in the adjacent
site controlled by Bob,  the state $|23\rangle_{f(b)}$  or $|14\rangle_{f(b)}$, or  in the non-adjacent
Bob mode, the state $|13\rangle_{f(b)}$  or $|24\rangle_{f(b)}$. Thus, to analyze  the role of the initial configuration of the system
into the building up of non-classical correlations, we just need to evaluate the time evolution of  the  quantum states  $|12\rangle_{f(b)}$,  $|23\rangle_{f(b)}$, and
$|13\rangle_{f(b)}$. As shown in the previous section, the latter can be calculated from the 
single-particle amplitude transitions given in Eq.~(\ref{singpart}).  In agreement with the results of Ref.~\cite{blumen},
these, after a straightforward calculation for the case of $N$=4,  can be 
expressed as:
%
%
%
\begin{eqnarray} \label{singn4}
\lambda_{k,k}(t)&=&\exp\left({-2i\gamma t}\right)\cos^2{\gamma t} \nonumber \\
\lambda_{k\pm 1,k}(t)=\lambda_{k, k\pm 1}(t)&=&\exp\left({-2i\gamma t}\right) i\sin{\gamma t}\cos{\gamma t} \nonumber \\
 \lambda_{k\pm 2, k}(t)=\lambda_{k, k\pm 2} (t)&=&-\exp\left({-2i\gamma t}\right)\sin^2{\gamma t}.
\end{eqnarray}
By inserting the above coefficients in the Eqs.~(\ref{boso2}) and (\ref{fer2}), and then by using the expressions~(\ref{matrice}) 
and (\ref{vonNeu2}), we can quantify  the building up of QC for the initial quantum states of interest. We find that
\begin{widetext}
\begin{align}
 E_P\left(|\psi^{(12)}\rangle_{f(b)}\right)_{A=\{1,2\}}=& \frac{1}{2}\sin^2(2\gamma\;t)\;\ln2  \nonumber \\
E_P\left(|\psi^{(13)}\rangle_{f(b)}\right)_{A=\{1,2\}}=& -\frac{[\cos(2\gamma\;t)-1]^2}{4}\ln\left[\frac{[\cos(2\gamma\;t)-1]^2}{2[1+\cos^2(2\gamma\;t)}\right] \nonumber \\
&-\frac{[\cos(2\gamma\;t)+1]^2}{4}\ln{\left[\frac{[\cos(2\gamma\;t)+1]^2}{2[1+\cos^2(2\gamma\;t)]}\right]} \nonumber \\
 E_P\left(|\psi^{(23)}\rangle_{f(b)}\right)_{A=\{1,2\}}=&0.
\end{align}
\end{widetext}
In Figure~\ref{fig1},  we report the time evolution of $E_P$, namely the amount the QC,
 for a two-fermion (-boson) system prepared in different  initial conditions. 
For the sake of simplicity, in this section we call time the adimensional parameter $\gamma t$.
In agreement with theoretical
predictions, at  $t$=0, $E_P$ is null, thus indicating the absence of  QC between the two particles localized in different nodes. After initial time,
$E_P$ remains  zero for the input state $|\psi^{(23)}\rangle_{f(b)}$,  while  it exhibits  oscillations, between 0 and $\ln 2$/2,  with a period of 
$\pi/2$ for the other two input states $|\psi^{(12)}\rangle_{f(b)}$ and $|\psi^{(13)}\rangle_{f(b)}$.  While in the former initial
configuration the dynamics of the two-particle  wavefunction does not result  into building up of non-classical correlations, in the latter 
cases the interference effects are able to produce  QC, whose periodic behavior is strictly related to  the evolution of the system 
in a ring lattice with a small number of nodes with periodic boundary conditions.  Here, the highest value of  $E_P$ is $\ln 2$/2, that is the half
of the maximum degree of quantum correlation achievable between   two particles, each one in a two-mode subsystem. 
It is worth noting that, for the partition considered, $E_P$  turns out  not to depend upon the quantum statistics of the particles involved in the process. Such a result is 
somehow unexpected due to the different number of states accessible  to the two systems during the time evolution. Indeed, unlike fermions, 
bosons can occupy simultaneously  the same node,  that is   the transition probability  to the states $|kk\rangle_b$'s  does not vanish as time  goes by.
Even if the time evolution of the bosonic wavefunction differs from the  fermionic one, for the specific partition here considered  the two-particle interference effects
lead to the same single-particle features, as, in particular, the reduced density matrices of the subsystems. As a consequence, also the
amount  of QC  is the same.

The results above reported are closely related to how the nodes of the ring lattice
are assigned to Alice and Bob. If we move from the partition $A=\{1,2\}$ and $B=\{3,4\}$
to the one $A=\{1,3\}$ and $B=\{2,4\}$,  not only the QC turn out to depend upon
the quantum statistics of the particles, but  the highest degree of correlation
can also be reached. Let us examine the time evolution of two bosons, or two fermions, 
initially occupying the nodes 1 and 3. After a straightforward calculation, we obtain
\begin{widetext}
\begin{equation}
E_P\left(|\psi^{(13)}\rangle_{f}\right)_{A=\{1,3\}}= \ln2\sin^2(2\gamma t)\qquad \textrm{and} \qquad E_P\left(|\psi^{(13)}\rangle_{b}\right)_{A=\{1,3\}}=0. 
\end{equation}
\end{widetext}

While in the two-boson system no correlation is created, in the two-fermion one $E_P$
exhibits periodic oscillations and reaches the maximum value,  namely $\ln{2}$, for $\gamma t=(2k+1)/(4 \pi)$ with $k\in \mathbb{N}$.

As  stated above, Bose-Einstein statistics allows for the localization of two bosons on the same site.  When the latter
is taken as initial configuration of the  two-particle system, the amount of non-classical correlations built up
does not change with time and remains equal to zero for any partition examined. This means that  time-evolved
states of the kind $|kk\rangle_b$  can always  be factorized in terms of an one-particle state in the  Alice modes and  one in the Bob modes.
Thus,  an initial bosonic bunching prevents  the appearance of QC in CTQW.

\subsection{Large number of nodes}

For a lattice with $N$ sites, the dimensions of the Hilbert space for the two-boson and two-fermion
systems are $N(N+1)/2$  and  $N(N-1)/2$, respectively. This makes  the analytical techniques inefficient for large $N$ both to 
solve the two-particle dynamics and to evaluate the degree of quantum correlation.  For this reason, we implemented a numerical approach that, once known the time evolution
of the state,  allows one   first  to diagonalize the  single-particle reduced density matrix given in Eq.~(\ref{matrice}),
and then  to estimate, at any time, $E_P$.  Such an approach is   used here to quantify 
the amount of quantum correlations stemming from   two-particle CTQW in an 1D ring lattice with $N=50, 70, 100$,
when Alice  accesses the first half of the lattice and Bob the second half (namely, the partition  $A=\{1\dots,N/2\}$,$B=\{N/2+1,\dots,N\}$).

 \begin{figure}[htpb]
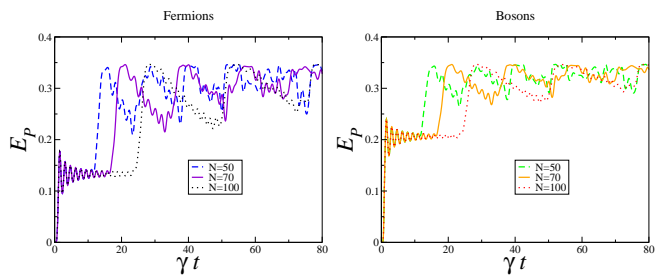

  \centering
  \includegraphics*[width=0.49\linewidth]{fig2.eps}
\includegraphics*[width=0.49\linewidth]{fig3.eps}            
 \caption{(Color online) Time evolution of  $E_P$ for the CTWQ of two fermions and two bosons,
 in the left and right panel, respectively, on a lattice with a number of sites $N=50$  (dashed line), $N=70$ (solid line), 
 and $N=100$ (dotted line). Both systems are initially prepared in the
 state $| \frac{N}{2} \frac{N}{2}+1\rangle_{f(b)}$. }
  \label{tantisiti}
\end{figure}

Fig.~\ref{tantisiti} displays the  time evolution of $E_P$ for two fermions (bosons) initially localized in two adjacent sites
belonging to different subsystems, corresponding to  the input state  $|N/2, N/2+1\rangle_{f(b)}$. As expected, at $t=0$ the amount of QC 
is zero.  As time goes by, $E_P$ increases and saturates, apart  from small oscillations, around  a specific value for a time interval $\gamma \tau$
which is found to be   linearly dependent upon the number $N$ of lattice nodes. These oscillations are closely related to the
two-particle interference effects stemming from the spreading of  the spatial wavepackets in a discrete lattice. In the limit
of large values of $N$,   the fluctuations disappear leading to a smoother behavior,
as it will be shown in the next section. After a time $\gamma\tau$, $E_P$ shows first  a rapid increase and then again large fluctuations
around a new saturation value.

\begin{figure}[h]
  \begin{centering} \begin{minipage}[c]{0.48\textwidth}
        \includegraphics*[width= 0.48\textwidth]{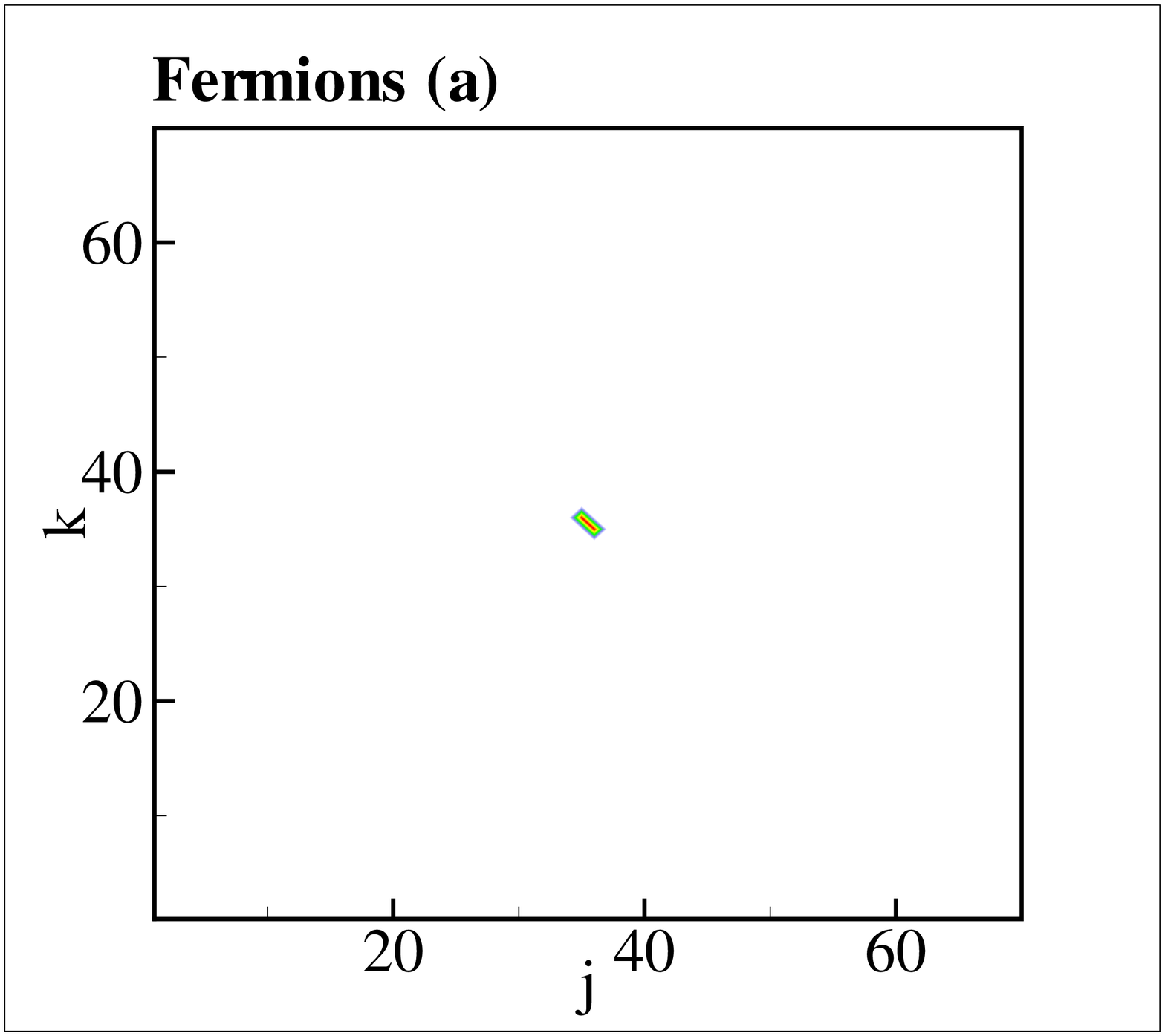}
       \includegraphics*[width= 0.48 \textwidth]{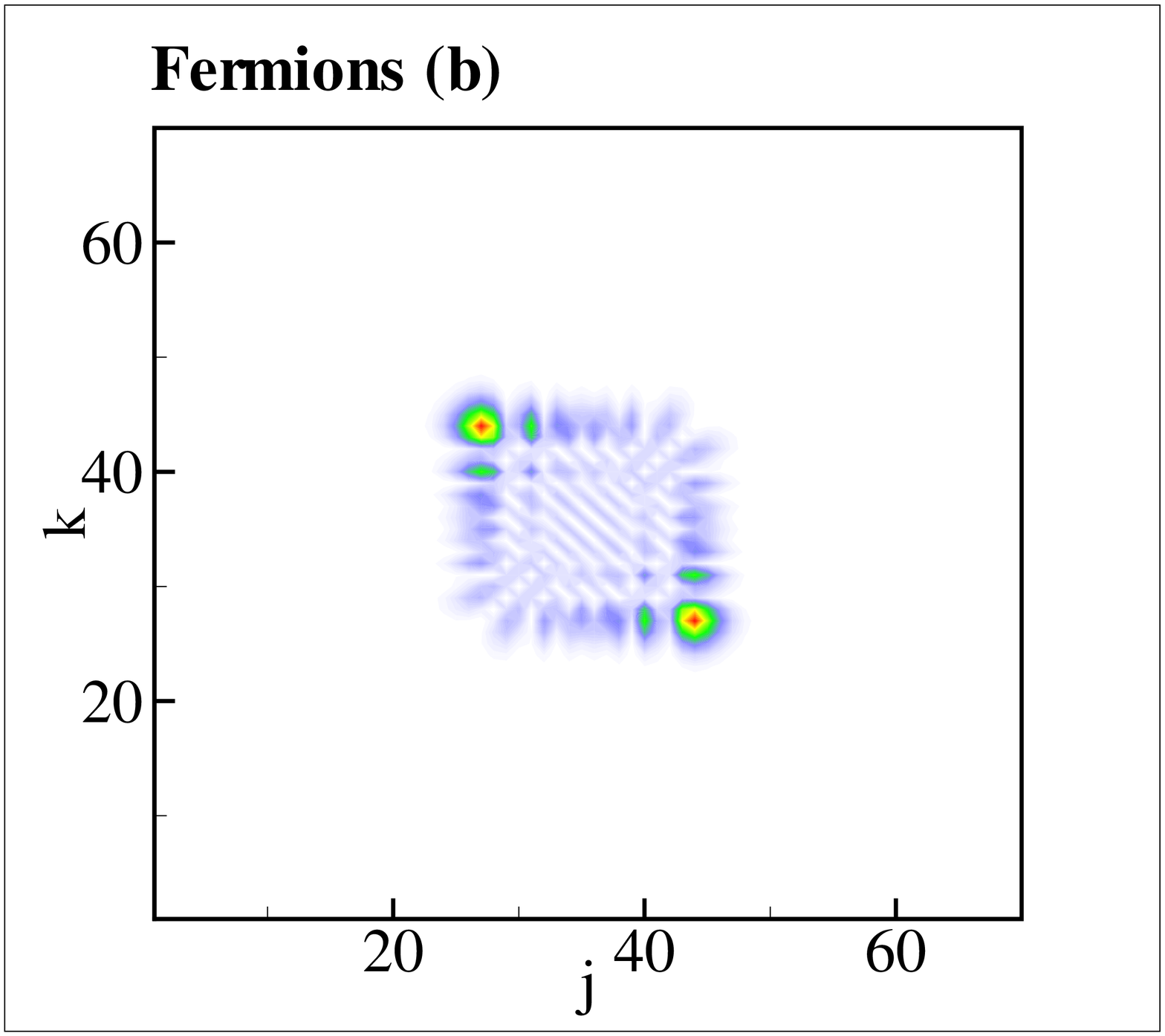}
        \includegraphics*[width= 0.48\textwidth]{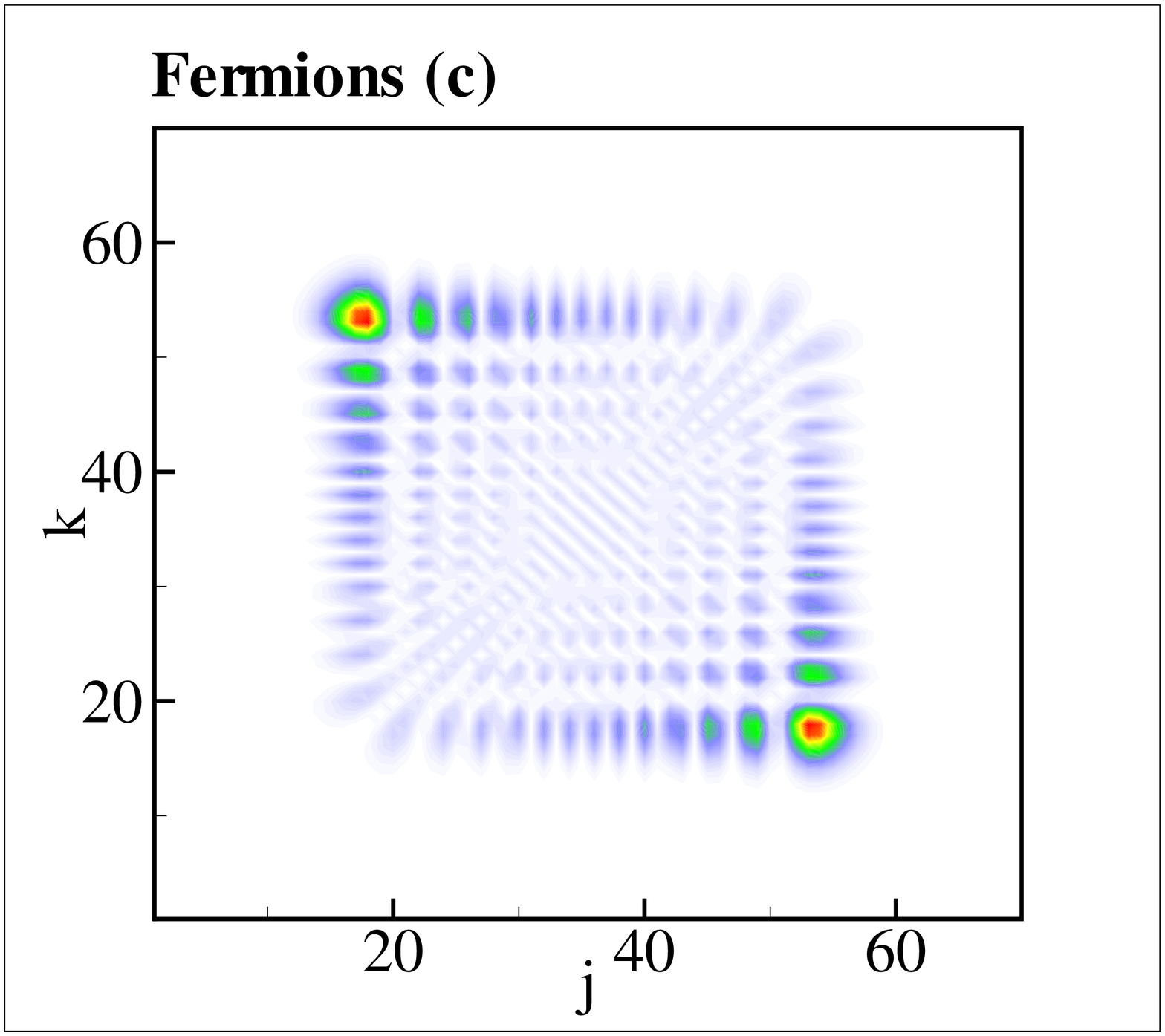}
        \includegraphics*[width= 0.48\textwidth]{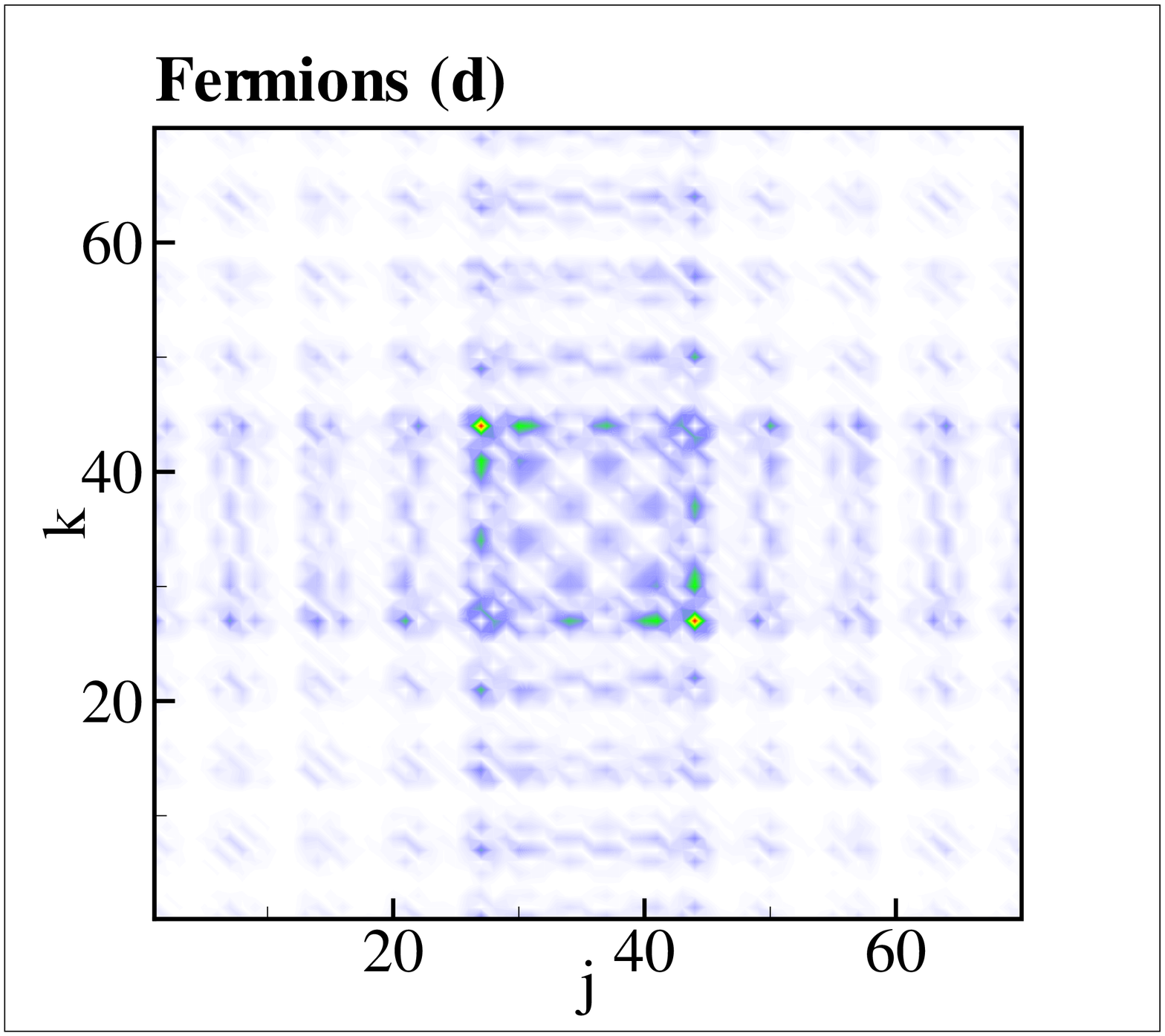}
    \end{minipage} 
  \begin{minipage}[c]{0.48\textwidth} 
      \includegraphics*[width= 0.48\textwidth]{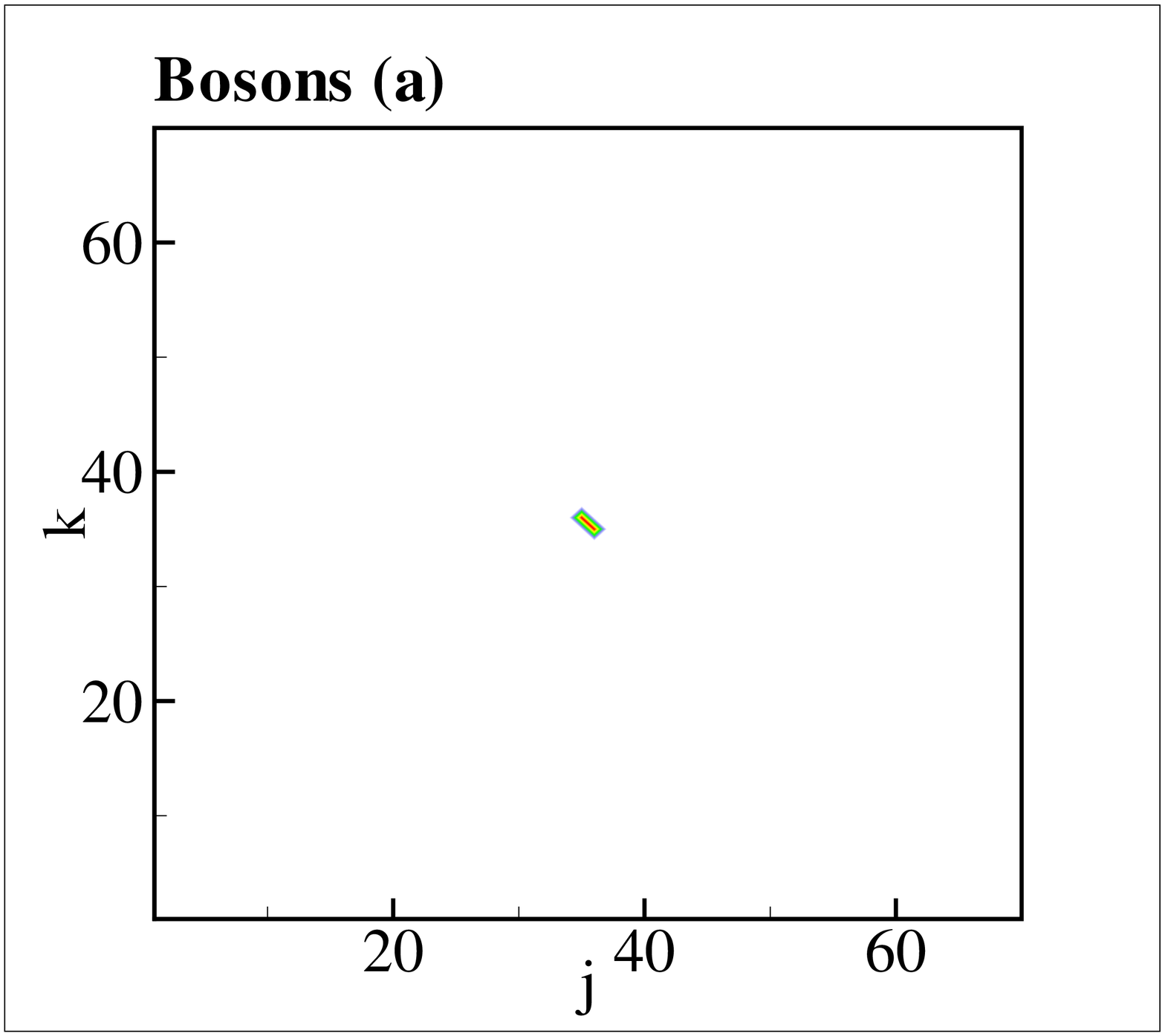}
       \includegraphics*[width= 0.48 \textwidth]{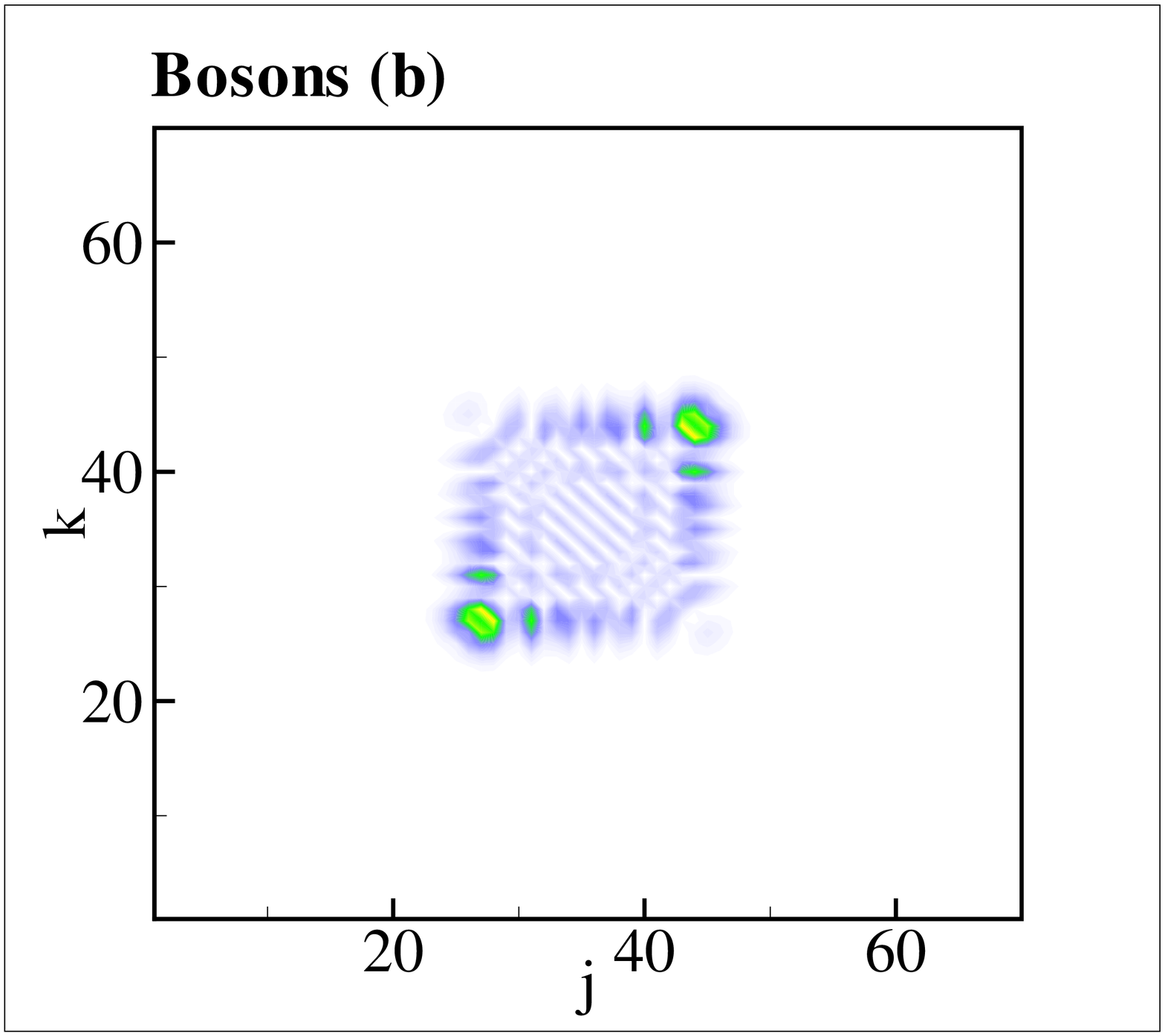}
        \includegraphics*[width= 0.48\textwidth]{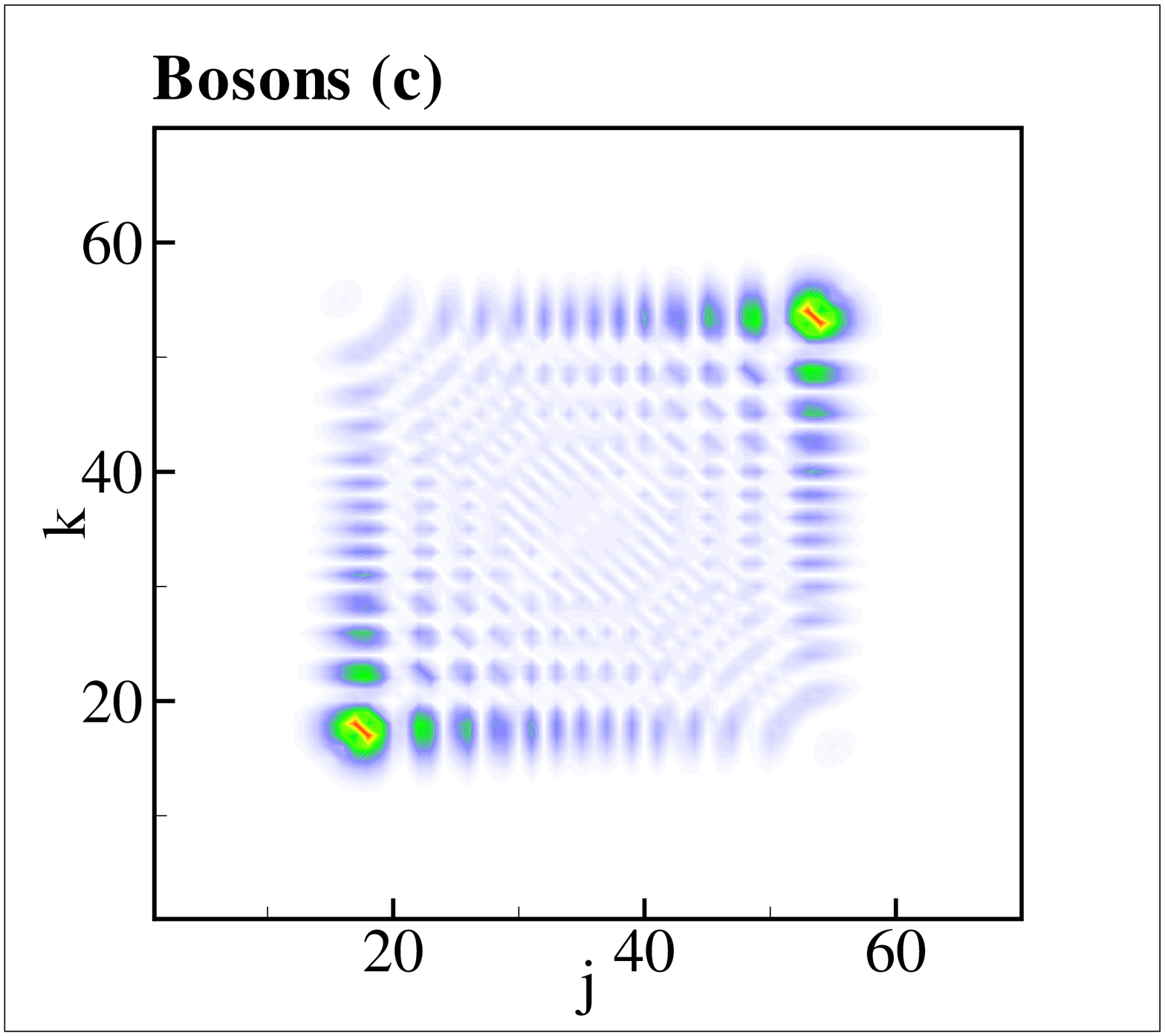}
        \includegraphics*[width= 0.48\textwidth]{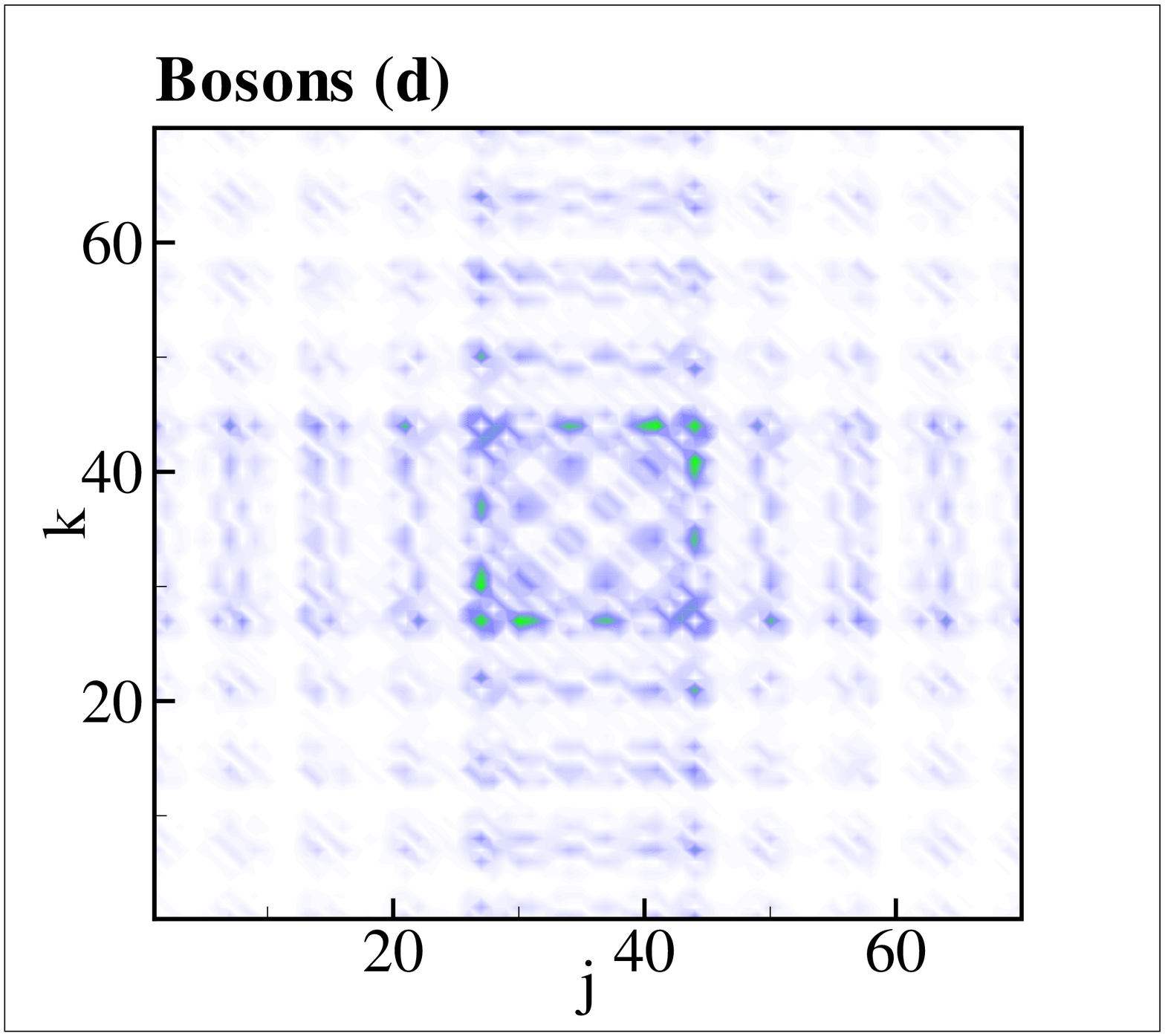}
    \end{minipage}\caption{\label{figpack} Left panels: two-particle correlation function $\Gamma_{kj}^{f}$
  of the two-fermion CTQW in a ring lattice with $N=70$ evaluated  at four different times: $\gamma t=0$, (a),
  $\gamma t=5$, (b), $\gamma t=10$, (c),  and $\gamma t=40$ (d). The input state is $| \frac{N}{2} \frac{N}{2}+1\rangle_{f}$.
  At short times, fermions exhibit ballistic evolution
  with spatial antibunching. At the final time,
  as a consequence of the interference stemming from the transmission through periodic boundary conditions, the two particle wavefunction
  is scattered along the whole lattice.  Right panels: two-particle correlation function $\Gamma_{kj}^{b}$
  of the two-boson CTQW in a ring lattice with $N=70$ evaluated  at four different times: $\gamma t=0$, (a),
  $\gamma t=5$, (b), $\gamma t=10$, (c),  and $\gamma t=40$ (d). The input state is $| \frac{N}{2} \frac{N}{2}+1\rangle_{b}$.
  At short times, bosons propagate ballistically along the lattice showing spatial bunching. After 
  reaching the boundary, the transmission through the  periodic boundary conditions induces the spread of
   two-particle
  wavefunction all over the lattice.}
  \end{centering}
\end{figure}

In order to get a better insight into  the appearance of QC, we focus on the time-evolution
of the two-fermion (-boson) correlation function $\Gamma_{kj}^{f(b)}(t)$~\cite{Bromberg},  namely the square modulus
of the coefficient $\mu_{kj, \frac{N}{2}  \frac{N}{2} +1}^{f(b)}(t)$. To this aim,
we report in Fig.~\ref{figpack}  the values of $\Gamma_{kj}^{f(b)}(t)$,
at four different times, for the case  of a  ring lattice with 70 nodes.
At initial time, the two fermions (bosons) are localized in two adjacent sites
(taken in the middle of the  lattice) in a separable state. After a relatively short time,
their wavepackets overlap and interfere thus
leading to the first rise of $E_P$. Then  the fermions (bosons) 
freely propagate along the lattices exhibiting spatial antibunching (bunching)~\cite{Lah},  as shown
by the two peaks positioned along the antidiagonal (diagonal) of the correlation matrix (see 
panels (b) and (c) of the Fig.~\ref{figpack}). During this ballistic evolution, the amount of QC
exhibits small fluctuations around a given value in the interval  $\gamma \tau$. 
The latter can now be viewed as a characteristic  parameter of the dynamics of the system. Indeed,
it corresponds to  the travelling time of the two particles along half a lattice before the transmission
thorugh periodic boundary conditions occurs. As expected, $\gamma \tau$ is independent
upon the quantum statistics of the particles. At longer times, the two-particle wavefunction 
is transmitted by the periodic boundary conditions, and their components interfere again 
creating  additional  QC. As reported in the panels (d) of 
Fig.~\ref{figpack}), at $\gamma t= 40$, $\Gamma_{kj}^{f(b)}$ reveals a checkered pattern
where both  fermions and bosons are scattered along the whole lattice and $E_P$
 oscillates in time.

\begin{figure}[h]
  \begin{center}
 \includegraphics*[width=0.6\linewidth]{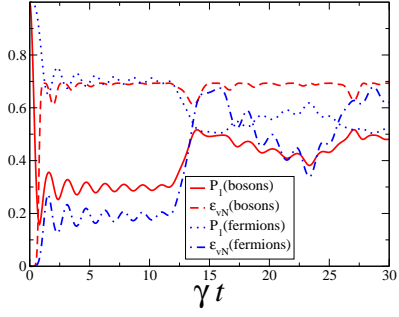}
    \caption{\label{entro}{ Probability $P_1$ of finding each particle in a different subsystem and von Neumann entropy $\epsilon_{vN}$,
    both for two bosons  and the for two fermions as a function of time for the case of a lattice ring with a number of nodes $N=50$.}}
    \end{center}
 \end{figure}

The time evolution of $E_P$ deserves further comments. 
The  results  reported here differ significantly from the ones found for  the case of a lattice with $N=4$. First,
  QC depend now upon the quantum statistics of the the particles considered, 
though the  qualitative behavior of $E_P$ of the two-fermion
 and -boson systems is the same.  Specifically, we find that  the degree of correlation between
 two bosons is always equal or greater than the one of the fermionic system. Such a behavior is at a first analysis counterintuitive .
 Unlike the two-fermion dynamics,  the bosonic  bunching lowers  the probability
  $P_1$ of  finding each of the two particles in a different subsystem (see Fig.~\ref{entro}).
  As  a consequence, according to Eq.~(\ref{entdue}),
  the amount of QC between two bosons
  should be lower than the one created in the fermionic system.
 However, also the degree of non-classical correlation of the two-particle state describing one particle
  in the Alice modes and the other one  in the Bob modes,  and quantified by  the von Neumann entropy,
  has to be   taken into account. In this case, the latter is larger for the
  two-boson CTQW and  the product $P_1\epsilon_{vN}$, namely $E_P$,
  turns out to be larger than the one corresponding to the two-fermion system.

Furthermore, we find  that the behavior in time  of $E_P$   is not    periodic. Indeed, as expected,
for lattices with a large number of sites, 
the transmission through the boundary conditions does not affect very much the time evolution of the two-particle wavefunction.
As a consequence  the interference among the different components of the wavefunction leads to a non-periodic  behavior of the correlations
Thus, the effect of the lattice periodicity on two-particle dynamics and the degree of quantum correlation
is lower for  lattices with higher values of $N$ and should  vanish in the limit
of a very large $N$.

\section{Two-particle free propagation in 1D space-continuous structures}\label{resultsII}

In this section, we numerically  study the building up of non-classical correlations in
 the free-propagation of two identical particles in a 1D  structure.

The physical model  here investigated consists of two non-interacting massive fermions (bosons)
travelling in a space-continuous structure. The latter can be viewed as
an infinite lattice with vanishing intersite distance without the interference effect stemming 
from the transmission through the periodic boundary conditions. In other terms, we are considering 
the continuous-space limit of the model examined in the previous section.
If  the free propagation of massive particles obeying the Bose-Einstein statistics constitutes
a helpful theoretical tool to investigate  the appearance of QC  due to the bunching
in a space-continuous structure, on the other hand  the fermionic system can also describe phenomena of physical interest,
like the electron diffusion in solid state systems.

The dynamics of the global system can be described by the two-particle Hamiltonian:
\begin{equation}\label{hcontin}
\mathcal{H}(x_1,x_2)=-\frac{\hbar^2}{2m}\left(\frac{\partial^2}{\partial x_1^2}+\frac{\partial^2}{\partial x_2^2}\right),
\end{equation}
with $m$ indicating the particle mass. Eq.~(\ref{hcontin}), that is the sum of two Laplacians,
is the continuous version of the Hamiltonian written in Eq.~(\ref{des}) for the CTQW of two
particles. Both of them are represented, at the initial time $t_0$, by a minimum-uncertainty
wavepacket,  that in the space representation, reads:
\begin{equation}
\phi_{\pm x_0,k}(x,t_0)=\left(\frac{1}{2\pi\sigma^2}\right)^{1/4}\exp{\left(ikx-\frac{(x\mp x_0)^2}{4\sigma^2}\right)} ,
\end{equation}
where $\pm x_0$ is the mean  position of the space wavepacket with variance $\sigma$,  and $k=\frac{\sqrt{2mE_{k}}}{\hbar}$
with $E_k$  kinetic energy of the particle. Thus the two-fermion (-boson) initial wavefunction $\Phi_{f(b)}(x_1,x_2)$ 
  can be written as:
  \begin{widetext}
\begin{equation}
\Phi_{f(b)}(x_1,x_2,t_0)=  \phi_{\overline{x}_0,k_1}(x_1,t_0) \phi_{-\overline{x}_0,k_2}(x_2,t_0) -(+)
  \phi_{\overline{x}_0,k_1}(x_2,t_0) \phi_{-\overline{x}_0,k_2}(x_1,t_0),
\end{equation}
\end{widetext}
with  the normalization condition $\int_{-\infty}^{+\infty} dx_1\int_{x_1}^{+\infty} dx_2 \left| \Phi_{b(f)}(x_1,x_2,t_0)\right|^2=1$.
The variance $\sigma$  of the two single-particle 
Gaussian wavepackets  and the distance $|2\overline{x}_0|$ between their centers
are such that,  at the initial time, their spatial overlap is practically null.

Here we estimate the amount of
QC between the spatial degrees of freedom of the two
particles. Specifically, we assume that the space domain $[-\infty, 0]$ is controlled by Alice,
while Bob controls to the spatial modes in the interval $[0,+\infty ]$. To quantify $E_P$ 
we adopt a numerical procedure. In fact, even if the dynamics of the system 
could be  evaluated analytically,  nevertheless
 the estimation of the  time evolution of $E_P$ as given in
 Eq.~(\ref{entdue})   would require involved calculations. 
 On the contrary, the implementation of  a numerical approach,
 exploiting the Crank-Nicholson finite difference scheme to
 solve the two-particle  time-dependent Schr\"odinger equation,
 permits to evaluate efficiently, at fixed time steps, the built up of the non-classical correlations.
 Here $E_P$ is evaluated by means of the linear entropy $\epsilon_L$, in place of the von Neumman entropy,
 in order to make   the numerical procedure less demanding, as shown in Ref.~\onlinecite{lentropy}.   
 Given the relation $\epsilon_L=1-\text{Tr}(\rho_A^2)$, $E_P$ reads
\begin{equation}
E_P=P_1(1-\text{Tr}(\rho_A^2))=P_1\left(1-\int_{-\infty}^0 dx \rho_A^2(x,x)\right),
\end{equation}
where the space-continuous nature of the model is explicitly taken into account. 
In the above expression, the probability $P_1$ of finding one particle in the Alice  side and the other in the Bob side,
is given by $\int_{-\infty}^0 dx_1 \int^{+\infty}_0 dx_2 |\Phi_{b(f)}(x_1,x_2,t)|^2$, while the
single-particle density matrix  describing the subsystem controlled by Alice
is $\rho_A(x,x^{\prime})=(1/P_1)\int^{+\infty}_0 dx^{\prime\prime} \Phi_{b(f)}(x,x^{\prime\prime} ,t)\Phi_{b(f)}^{\ast}(x^{\prime},x^{\prime\prime},t)$
with $x,x^{\prime}\in [-\infty, 0]$.

We examine the model for  two different initial conditions: particles 
with the same velocity, that is $k_1$=$k_2$,  and particles with opposite
velocities, that is $k_1$=-$k_2$.  The latter case mimics a 
collision event between two  non interacting identical particles.

\begin{figure}[h]
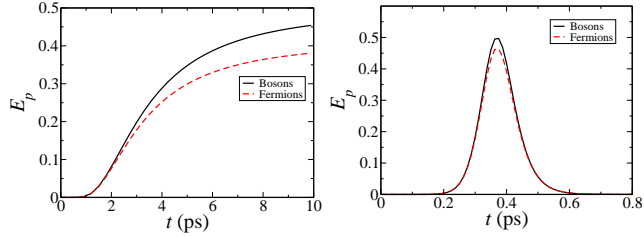

  \begin{center}
    \begin{minipage}[c]{0.48\linewidth} 
      \includegraphics*[width=\linewidth]{fig13.eps}
    \end{minipage}
    \begin{minipage}[c]{0.48\linewidth} 
      \includegraphics*[width=\linewidth]{fig14.eps}
    \end{minipage}
    \caption{\label{continuo} (Color online) Left panel: 
     $E_P$ as a function of the time for two fermions (dashed line) and two bosons (solid line) 
       moving with the same velocity.
     Such a condition is equivalent to the one where the particles have
     zero initial kinetic energy. In order to make the numerical
     implementation of wavefunction dynamics simple, here we 
     considered  $E_{k}=0$. The particles are initially described  by two wavepackets
     with initial variance $\sigma$=5nm and $\overline{x}_0$=20 nm.
    Right panel: 
     $E_P$ as a function of the time for two fermions (dashed line) and two bosons (solid line)  moving in opposite directions.
     The particles have the same kinetic energy $E_{k}=10$ meV,  and, at the initial time,
      are also described by two wavepackets
     with variance $\sigma$=5nm and $\overline{x}_0$=20 nm.
     In all  the numerical calculations we take $m$=9.1$\times$10$^{-31}$Kg.}
  \end{center}
\end{figure}

The former case can be thought as the space-continuous analogous of  the CTQW on a site lattice
with two particles initially localized in specific sites. At the initial time,  both Alice and Bob 
have a particle:  the centers of the two single-particle wavefunctions are located in different spatial
subdomains and the space overlap between the wavefunctions is zero. This
makes  the degree of non-classical correlation vanishing, as shown in left panel of Fig.~\ref{continuo}.
As time increases, single-particle wavepackets spread out making 
the spatial overlap between them not negligible anymore. Now the probability
amplitudes of  the two-fermion (-boson) wavefunction $\Phi_{b(f)}(x_1,x_2)$ 
can interfere and quantum correlation builds up. The latter increases with time 
ad finally reaches a stationary value depending  upon the quantum statistics of the particles involved:
it is higher for bosons. Apart from the absence of oscillations which can be related to the space-continuous structure, 
the time evolution of $E_P$ of  the fermionic (bosonic) system is,  at least at short times, 
in qualitative agreement with the dynamics of the quantum correlations
appearing in the model investigated in the previous section. At longer times,
due to the different boundary conditions of the two systems, 
a discrepancy between the two entanglement-time behavior is observed. While in the  ring lattice
the two-particle wavefunction is transmitted  through periodic boundary conditions  and this results in a further increase
of the amount of quantum correlations,  the space-continuous structure is supposed to be infinite and the  wavepackets can only spread 
with no additional interference effect due to transmitted components.

For the case of   two fermions (bosons) running in opposite directions,
QC
exhibit a peculiar behavior (see the right panel of Fig.~\ref{continuo}). $E_P$
increases while the two fermions (bosons) are approaching each other. Specifically,
when the centers of the two wave packets reach the minimum distance,
 two-particle interference gives
the maximum amount  of quantum correlations. Finally, $E_P$  drops again to  zero
once the particles get far apart and the corresponding Gaussian wave packets 
exhibit a negligible spatial overlap.  This behavior is in agreement 
with previous analyses, adopting other entanglement criteria,
of the dynamics of  the QC stemming from 
carrier-carrier scattering events in semiconductor nanostructures~\cite{bbb,berreg}. 

\section{Conclusions}\label{conclusions}
Recent experiments addressed the appearance of non-classical correlations
in  CTWQ  of photon pairs in coupled waveguide lattices~\cite{Bromberg, Peruzzo}. Such correlations 
have been related to non trivial interference effects due to the quantum
statistics of the particles. In this perspective, an accurate analysis of  the quantum correlation created
in few-particle CTQW models is undoubtedly of great  interest  given their 
experimental feasibility and  potential application in quantum communication and quantum computation theory~\cite{Bromberg, Peruzzo}. 

Here, we investigated  the appearance of QC in  CTWQ of 
two non-interacting bosons (fermions) on an 1D ring lattice with periodic boundary conditions.
Specifically, in our model the topology of the graphs examined is very simple,
that  is each node is connected to its first neighbours and this permits to express
the Hamiltonian, ruling the two-particle dynamics,  in terms of the so-called discrete single-particle  Laplacian.
Indeed, our CTQW well describes  the free propagation of two non-interacting identical particles in a periodic system.
Given the key role played by the exchange symmetry  into  the emergence of non-classical
correlations, their quantitative evaluation in our physical system required  the use
of a suitable  criterion  which takes into account  the indistinguishability of the particles. 
To this purpose  we adopted the Wiseman and Vaccaro
approach~\cite{wiseman} which allowed estimating the degree of quantum correlation between two parties of the system
(each possessing one particle and accessing to a given set of nodes) with no violation of the particle local number 
superselection rules, as occurring in the preparation,
manipulation and measurement of the experimental implementations of   CTQW.

In agreement with the theoretical predictions~\cite{Bromberg, Peruzzo},  results indicate that the building up of QC
 in our system is  due to the two-particle interference effects between the propagating wavepackets. 
Indeed, the time evolution of the quantum system affects the  degree of correlation. Specifically,
we find that, for  CTQW on discrete ring lattices with a small number of nodes,
the   transition amplitudes depend sinusoidally on time due to the transmission of the wavepackets
through periodic boundaries conditions. As a consequence,  also QC exhibit a periodic-time behavior.
They are also affected by the quantum statistics of the particles involved in the process and by the partition of the system. Specifically,
the production of the maximum degree of non-classical correlation 
between subsystem accessing  non adjacent nodes occurs, cyclically, 
only for   two-fermion systems initially prepared in a suitable input state.  Such a result seems
 consistent with experiments in electron  Hanbury Brown-Twiss interferometers
showing the emergence of QC between two identical charge carriers   at
couples of non adjacent drains~\cite{Neder,Samue}.  For  two-particle CTQW on 1D discrete ring lattice 
with a large number of nodes, the time evolution of the quantum system is less affected 
by transmission through periodic boundary conditions and this leads to the disappearance 
of the cyclical behavior of the QC.  In particular, the  latter
are still related to the interference stemming from the overlap of the wavepackets, but 
they turn out to be almost insensitive to  the quantum statistics of the particles,  a part from small fluctuations related to the discrete nature of the ring lattice.

Finally,  we analyzed the building up of QC in the free-propagation
of two identical particles in an 1D structure. Such a system can be considered  the continuous 
space limit of an 1D ring lattice  with a large number of sites and vanishing intersite distance.
It provides an useful guideline both to validate  the results found for the discrete case
and to  compare our outcomes  with the analyses of non-classical correlations appearing in scattering 
events in 1D semiconductor structures~\cite{bbb,berreg}. Two different initial conditions have been  examined: 
particles moving with the same velocity in the same or in the opposite direction.
In the former case, we find that  
the QC of the system first show an initial increase due to the interference of the spreading single-particle wavepackets 
and then, for sufficiently long times, reach a stationary value, different from zero, depending upon quantum statistics. 
Such behavior is  in qualitative agreement with the one found for  CTQW in 1D ring lattice
(apart  from the absence of the effects  related to the transmission through  periodic boundary conditions
and to the discreetness of the lattice). On the other hand, for the case of particles propagating one against the other, as the 
fermions (bosons) get closer, the degree of quantum correlation increases and attains its maximum values
in correspondence of the maximum spatial overlap of the single-particle wavepackets.
When  the two particles get away from each other,  the amount of QC decreases until it vanishes.
Such a behavior appears to be consistent with the time evolution of the entanglement created
in binary collisions of electrons in semiconductor structures~\cite{bbb,berreg}.

Despite  the simplicity of the graphs here investigated, it seems   reasonable
to assume that the results obtained are representative, at least qualitatively, of the behavior
of the QC between two  indistinguishable particles
in more complex structures, such as 2D lattices or graph with larger connectivity. Thus,
the analysis here reported  represents a valuable starting point to investigate the 
emergence of non-classical correlations in  other models of  or fermionic    or bosonic CTWQ,
where also particle-particle interaction could be considered. Furthermore, it would be
of interest to insert in such models environmental noise such as the one due to lattice disorder. 
Indeed, the latter can be viewed as a decoherence source and, as a consequence, should
result into peculiar phenomena, among which the sudden death or the revival
of quantum correlations.

\end{document}